\documentclass[3pt,onecolumn]{article}
\usepackage{graphicx}

\usepackage{color}

\newcommand{\ben}{\begin{enumerate}}
\newcommand{\een}{\end{enumerate}}

\newcommand{\be}{\begin{equation}}
\newcommand{\ee}{\end{equation}}
\newcommand{\bea}{\begin{eqnarray}}
\newcommand{\eea}{\end{eqnarray}}

\begin{document}
\begin{flushright}
\end{flushright}
\vspace{0.1cm}
\thispagestyle{empty}

\begin{center}
{\Large\bf 
Duality of maximal-torsion and close-packing in double helices
}\\[13mm]
{\rm  Kasper W. Olsen\footnote{kaspeolsen@math.ku.dk}\\
Department of Mathematical Sciences, University of Copenhagen\\ 
Universitetsparken 5, 2100 Copenhagen \O, Denmark\\
\vspace{5mm}
Jakob Bohr\footnote{jabo@nanotech.dtu.dk}\\
DTU Nanotech, Technical University of Denmark\\
{\O}rsteds Plads, 2800 Kongens Lyngby, Denmark}\\[18mm]

\begin{abstract}
It is suggested that there is an interesting transition in double helices. The transition is between a helix, which has maximal torsion for the constituting helical strands, and a double helix which is nearly optimally packed illustrated by corresponding dual points on the rope curve. The possible physical relevance for denaturation and renaturation of DNA is discussed. 
\end{abstract}
\end{center}

\section{Introduction}
Essentially, there are three distinct mechanisms for holding together a double helical structure (a rope, DNA, etc...).
Double helices can be held together by forces which are tensile in origin, i.e. geometrically expressed as fixed boundary conditions in combination with maximising  the length. This leads to a solution on the upper branch of the "rope curve". The rope curve is the generic curve that shows the total length of a double helix as a function of the number of turns \cite{bohr2011a}. Another type of interactions that hold a double helix together is molecular forces which typically favours a relatively even distribution of electronic density and hence close-packing of the strands. This condition appears on the lower branch of the rope curve. The close-packed structure is the idealised "natured" state \cite{olsen2009}. A third mechanism not considered here is through capillary forces, i.e. through a thin fluid layer between the strands.

In this letter, we discuss an effect which describes a transition between the first and the second type of helices on the rope curve. The rope curve defines the limiting cases where the two strands are in direct contact, while the inner points defines all "denatured" states, i.e. where the strands are not in contact. Denatured states with the same amount of twist lies on a vertical line. For all practical purposes, the close-packed structure and a maximal-torsion structure are on the same vertical line, and are therefore connected by denaturing and renaturing through constant twist.

The thermodynamics of bubble denaturing, i.e. local opening, of DNA has been subject to various studies, e.g. \cite{hanke2003,hanke2008,hanke2013,peyrard2004,cocco1999}. 
Bubble denaturing of DNA studied in fluid channel systems are free to rotate with no fixtures of the ends, see e.g. \cite{marie2013}. In biological systems local denaturing can occur without the presence of linking-number changing enzymes, leading to rotationally fixed ends. In a simplified form, this is what is studied in the present letter. 

\section{Method}
Our starting point is helices modelled as flexible tubes with hard walls, and work on the packing of such helices is e.g. \cite{pieranski1998,gonzalez1999,maritan2000,stasiak2000,przybyl2001,neukirch2002,bruss2012}.
Consider a helix made from circular strands of length $L_M$ and diameter $D$ (any number of strands, $N$, will do). The length of the helix is $H_M=L_M \sin v_\bot$, where $v_\bot$ is the pitch angle, and its total twist is
$\Theta_M=L_M \cos v_\bot/a$. 
Geometrically, the pitch angle is the angle between a tangent of the helix and the horizontal plane. 
The pitch angle is determined by the relation,  $\tan v_\bot = h/a$, where $h$ is the reduced helical pitch and $a$ is the helical cylinder radius. Therefore the total twist can be written as
\begin{equation}
\Theta_M = L_M \frac{\sin v_\bot}{h}\, .
\end{equation}
The total torsion is the integral
\begin{equation}
\Omega_M = \int_0^{L_M}\tau(s) ds = \frac{h}{a^2+h^2}L_M = L_M \frac{(\sin v_\bot)^2}{h}
\end{equation}
I.e. 
\begin{equation}
\Omega_M = \sin v_\bot \Theta_M 
\end{equation}
Imagine an experiment where $H_M$ is changed by pulling. Then, the length changes by an amount,
\begin{equation}
dH_M = L_M \cos v_\bot dv_\bot,
\end{equation}
and the torsion changes by
\begin{equation}
d\Omega_M = \Theta_M\cos v_\bot dv_\bot + \sin v_\bot d\Theta_M,
\end{equation}
so that
 \begin{equation}
\frac{d\Omega_M}{dH_M} = \frac{\Theta_M}{L_M}+\sin v_\bot \frac{d\Theta_M}{dH_M}\, .
\end{equation}
The scale-invariant quantity obtained by multiplying this expression with the radius of the tubes, i.e.
\begin{equation}
\frac{D}{2}\frac{d\Omega_M}{dH_M} = \frac{D}{2a} \frac{\cos 2v_\bot}{\cos v_\bot}
+\sin v_\bot \frac{d}{d v_\bot}(\frac{D}{2a}),
\end{equation}
is evidently a function of $v_\bot$ that measures how the torsion changes under strain. We define
the {\it differential torsion} as,
\begin{equation}
f_\Omega (v_\bot)= \frac{D}{2}\frac{d\Omega_M}{dH_M} \, .
\end{equation}
Similarly, the {\it differential twist}, $f_\Theta$, is defined as,
\begin{equation}
f_\Theta (v_\bot)= \frac{D}{2}\frac{d\Theta_M}{dH_M} \, ,
\end{equation}
and its significance for the strain-twist coupling of double helices is discussed in \cite{olsen2011}.

\section{Results}
Consider the double helix, i.e. $N=2$.
The differential twist, $f_\Theta$, is zero at $v_\bot=39.4^\circ$ (Fig. 1A), which was observed in \cite{olsen2011}, and this specific angle has earlier been denoted the zero-twist angle. The differential torsion, $f_\Omega$, is maximal at about $v_\bot=29^\circ$, and is zero at the non-trivial value $v_\bot=45^\circ$ (Fig. 1B). 
For a double helix, the configuration with $v_\bot=45^\circ$ has been designated as the tightly-packed double helix \cite{olsen2009}. It is also an angle, where the strand has maximal torsion, and we therefore identify it as the maximal-torsion angle (labeled MT). The maximal-torsion angle corresponding to $f_\Omega =0$ can be shown to approach $55^\circ$ for large $N$.
\begin{figure}[h]\centering
\caption{\it (A) Differential twist, $f_\Theta$, for a double helix as a function of pitch angle $v_\bot$. (B) Differential torsion, $f_\Theta$, for a double helix as a function of pitch angle $v_\bot$.}
\includegraphics[width=6.0cm]{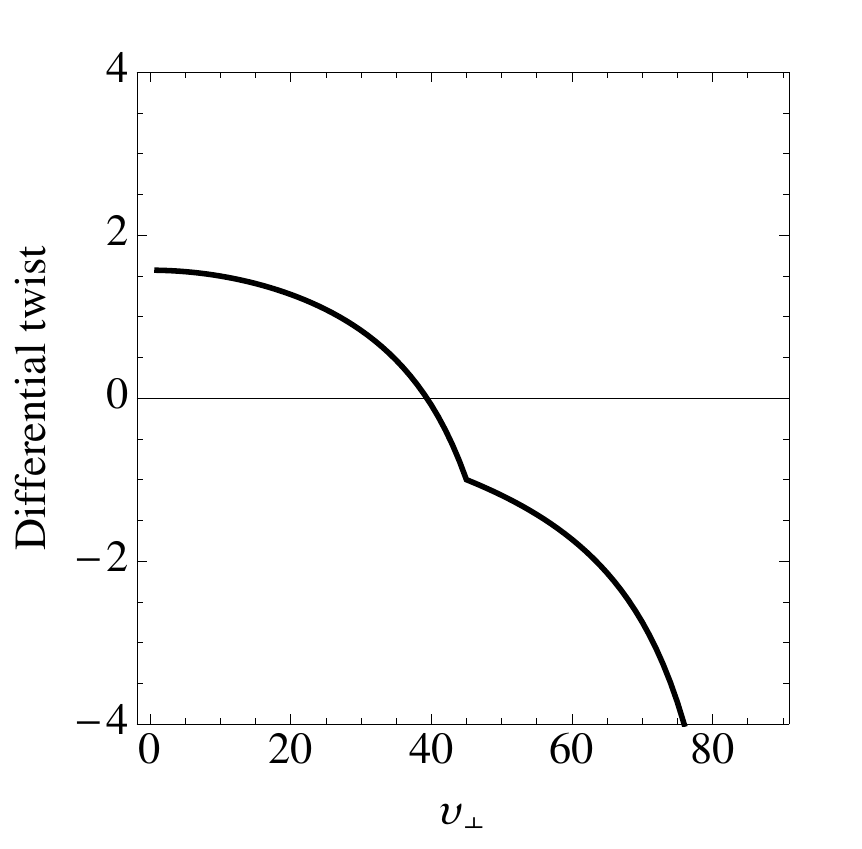}
\includegraphics[width=6.0cm]{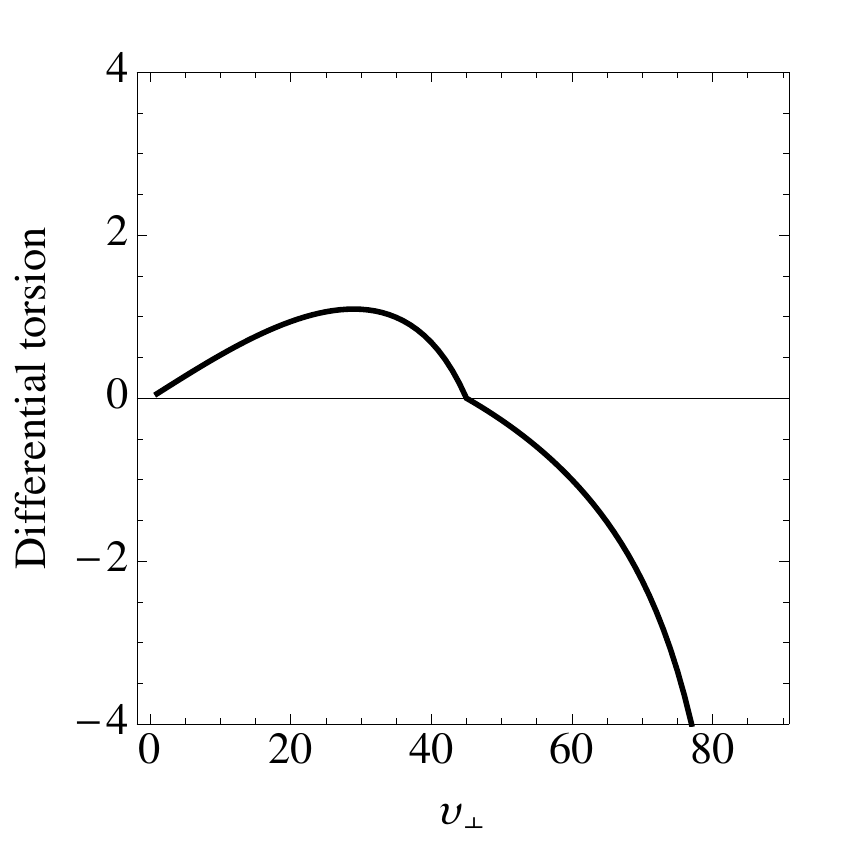}
\end{figure}

If we assume, that the number of turns, $n=\Theta_M/2\pi$, in the double helix is kept constant, we can imagine a situation where there is a transition between states having the same number of turns. 
The transition under conserved twist is shown on the rope curve as taking place between two "dual" points. On the rope curve, these are points defined to be at opposite locations in the vertical direction, see Fig. 2.

For a double helix with maximal torsion, the dual configuration is found numerically to have a pitch angle of $v_\bot= 32.9^\circ$. 
The $v_\bot=32.9^\circ$ geometry (labeled CP$^*$) is virtually identical to the actual close-packed structure at $v_\bot=32.5^\circ$ (labeled CP). The close-packed structure optimises a volume fraction \cite{olsen2009} defined as the volume of the two strands divided by the volume of a smallest cylinder enclosing them,
i.e. $f_V=2V_S/V_E$ 
(the volume fraction at the CP structure is 0.7694, at the CP$^*$ structure it is 0.7693, and at the MT structure it is 0.7071). 
Conversely, the dual state of this close-packed structure is a state with pitch angle $v_\bot = 45.3^\circ$.
{\it We conclude, that for any symmetric double helix, there is a near perfect duality between the maximal-torsion and the close-packed state}.

 It should be noted that the maximal-torsion geometry has no inner channel, while its dual CP configuration does have a channel, as discussed in detail in \cite{olsen2009}. This happens at the "expense" of being a much shorter configuration: 
The corresponding factor is $0.768$, see table 1.
\begin{figure}[h]\centering
\caption{\it Rope curve: extension $H$ of double helix as a function of the number of turns, $n$. For convenience, the strand length is chosen to be $L=1$, while the strand diameter is $D=1/100$. The upper part of the curve is obtained by twisting two straight strands together under strain, this is where the maximal-torsion (MT) geometry is located. The lower part of the curve is obtained by twisting the strands together on a large imaginary cylinder whose radius is successively becoming smaller.  This is where the close-packed (CP) structure is located. The apex corresponds to the zero-twist structure.}
\includegraphics[width=6.0cm]{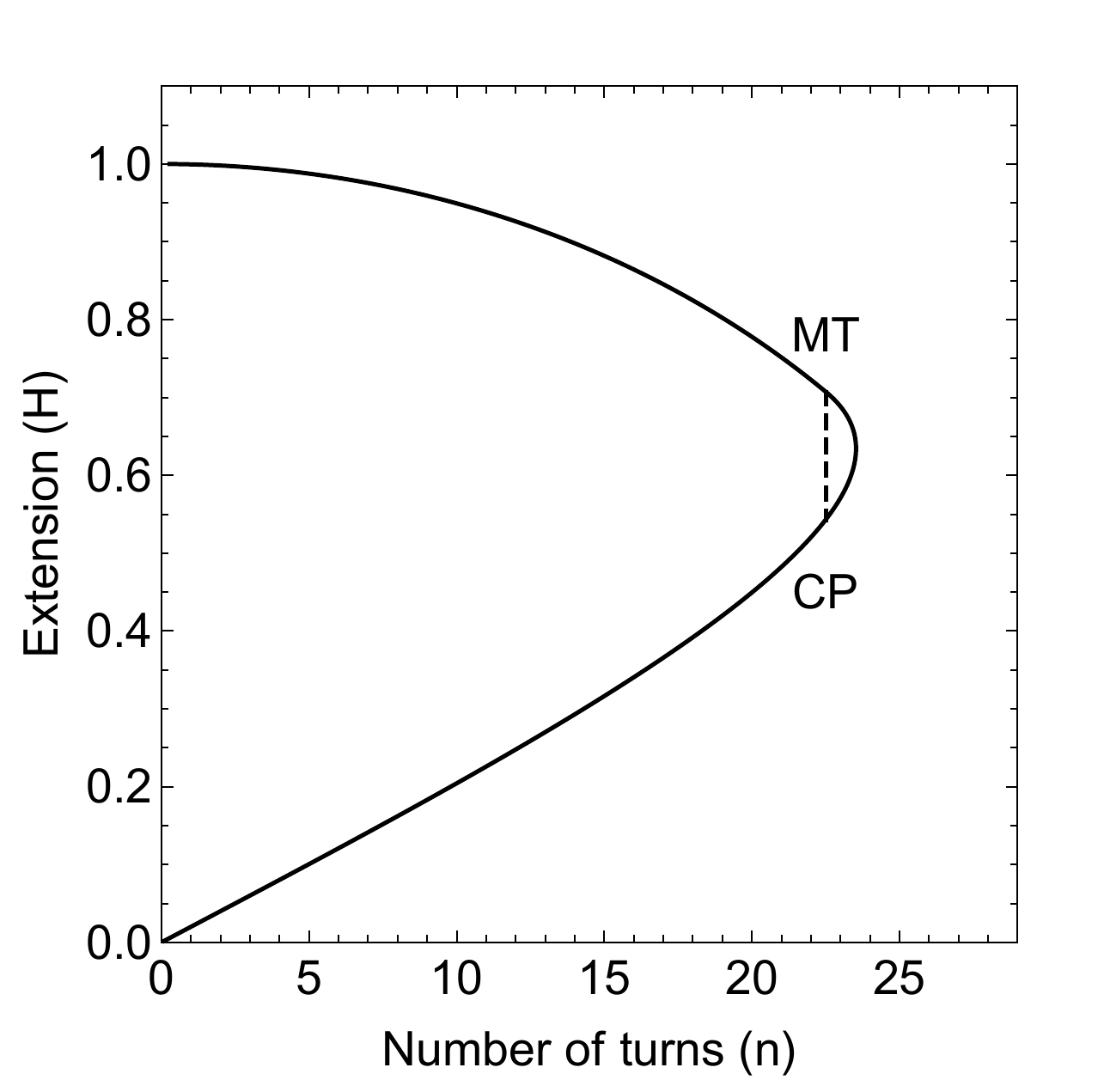}
\hspace{15mm}
\includegraphics[height=6.0cm]{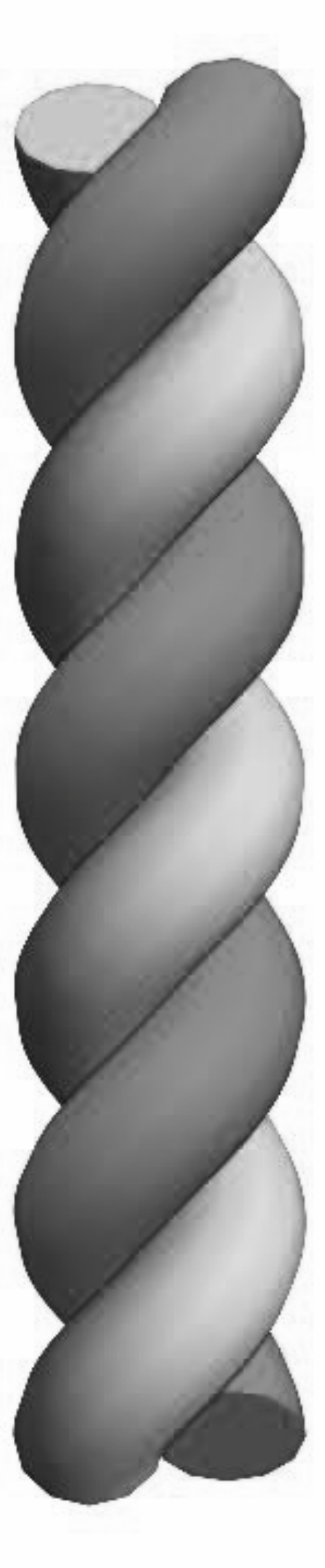}MT
\hspace{10mm}
\includegraphics[height=5.0cm]{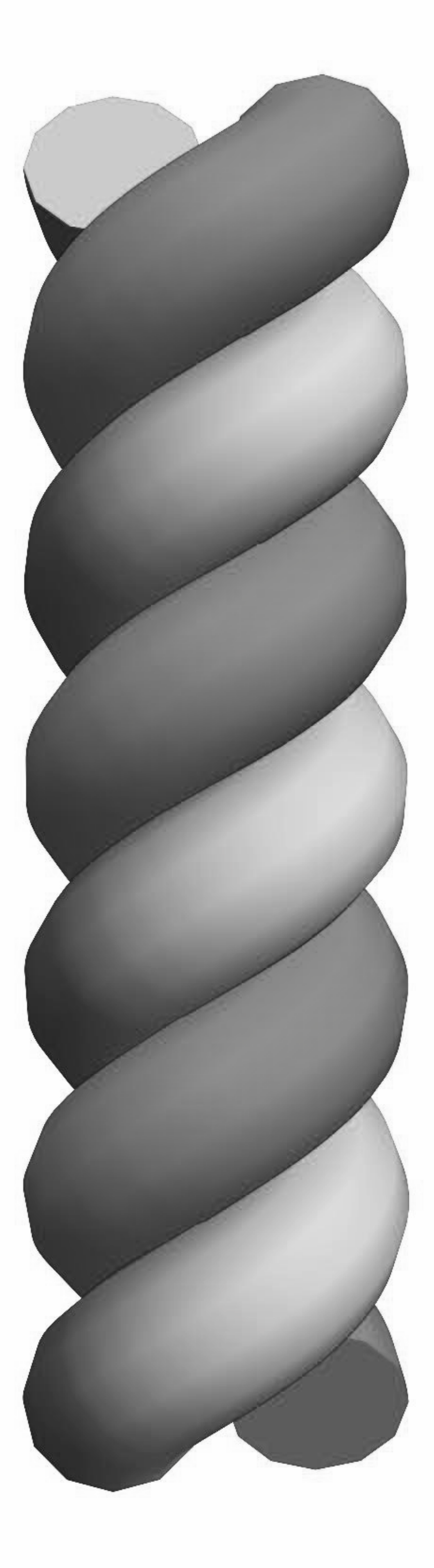}CP
\end{figure}

\begin{table}[h]
\caption{\it Maximum-twist and maximum-torsion structures for double helix. 
The "dual angle" is the opposite point on the rope curve with the same number of turns as the maximum torsion angle. The "length factor" is the fraction between the length of these two geometrical structures (MT/CP) while the length of the individual strands is conserved.}
\begin{center}
\begin{tabular}{|c|c|c|c|c|}
\hline
$N$ & $f_\Theta=0$  & $f_\Omega = 0$ & Dual angle & Length factor\\
\hline
2 & $39.4^\circ$ & $45.0^\circ$ & $32.9^\circ$ & $0.768$\\
\hline
\end{tabular}
\end{center}
\label{default}
\end{table}

\section{Discussion}
Total torsion and total twist are two geometrical quantities which can be computed for space-curves.
Their behaviour under strain lead us to define the functions, $f_\Omega$ and $f_\Theta$.
The condition $f_\Theta=0$ defines the zero-twist structures, which have been studied before 
in the context of overwinding of DNA, and rope-making (see \cite{bohr2011a,olsen2011}). The condition $f_\Omega = 0$ is apparently new.
For a space curve, the two quantities are the same assuming that the Frenet-Serret frame is used. 
For a space curve on a ribbon, or surface, the two quantities are distinct. For the double helix, it is two different curves, namely the centreline on a helicoid surface, and the space curve (i.e. helix) defining one of the two strands.

Maximumly twisting both strands collectively in a double helix leads to the zero-twist structure, or the apex of the rope curve. Maximumly twisting an individual strand leads to the maximal-torsion solution at $45^\circ$.
As shown above, approaching the maximal torsion configuration is a way to attain the dual of the close-packed structure. Hence, one can wonder whether this duality principle has been utilised in Nature? 

An obvious molecular candidate is that of DNA, more specifically in relation to replication and transcription. In \cite{olsen2009} it was shown that the geometry of double-stranded DNA is determined by close-packing, i.e. that it optimises a volume fraction and is a CP structure to a very good approximation. 
Initially, the denaturation of double-stranded DNA (dsDNA) takes place in isolated bubbles along the molecule 
\cite{hanke2003,hwa2003}. A mechanism that maintains the twist over long distances is therefore advantageous. To a large extend this mechanism avoids the presence of enzymes that modify the linking-number of DNA.
It is not clear that the maximal-torsion state can be reached by dsDNA in vivo as the 
full effect of the base-pairs \cite{manghi2016} is not described in this letter.
If the maximal-torsion state can be reached, we can write the denaturation and renaturation of DNA as a reaction equation, i.e.
\begin{equation}
dsDNA \longleftrightarrow dsMT 
\longleftrightarrow ssDNA \, ,
\end{equation}
where MT is the maximal-torsion state, and ssDNA implies single-stranded DNA. This should be contrasted with the usual picture, i.e.
\begin{equation}
dsDNA 
\longleftrightarrow ssDNA \, .
\end{equation}

Perhaps biochemical studies, as for example Raman spectroscopy \cite{barhoumi2008}, can detect the difference between a double-stranded maximal-torsion state, or two separated strands.
It remains to be seen whether the maximal-torsion state plays a role in denaturing/renaturing of DNA. Even if the MT state is not fully expressed in dsDNA, there is one denaturing path that is more advantageous than any other. 

Another area, where the suggested duality might be at play, is within DNA supercoils \cite{irobalieva2015}. Often DNA finds itself in a supercoiled structure, i.e. for example plectonemic DNA. Such superstructures, which are effectively a double helix, can be subject to the same kind of analysis but generally has much larger pitch angles than the maximal-torsion structure.


\end{document}